\documentclass{PoS}
\newcommand{\dd}{{\mathrm{d}}}
\newcommand{\Z}{{\mathbf{Z}}}

\def\di{\displaystyle}

\def\bg{\begin{eqnarray}\begin{array}{rcl}\displaystyle}
\def\eg{\end{array} &\di    &\di   \end{eqnarray}}
\def\bm#1{\begin{eqnarray}\begin{array}{#1}\di}
\def\bmo#1{\begin{eqnarray*}\begin{array}{#1}\di}
\def\bml#1#2{\begin{eqnarray}\begin{array}{#1}\label{#2}\di}
\def\bgo{\begin{eqnarray*}\begin{array}{rcl}\displaystyle}
\def\ego{\end{array} &\di    &\di \nonumber  \end{eqnarray*}}

\def\btensor#1#2{\left#1\begin{array}{#2}\di}
\def\brtensor#1#2#3{\ren#3\left#1\begin{array}{#2}}
\def\botensor#1#2{\renew\left#1\begin{array}{#2}}
\def\etensor#1{\end{array}\right#1}

\def\eq#1{(\ref{#1})}
\def\Eq#1{Eq.~(\ref{#1})}

\def\tr{{\rm tr}}

\def\s0#1#2{\mbox{\small{$ \frac{#1}{#2} $}}}
\def\0#1#2{\frac{#1}{#2}}

\newcommand{\nn}{\nonumber}

\title{CP invariance of chiral gauge theories and Majorana-Yukawa
couplings on the lattice}

\ShortTitle{CP invariance of chiral gauge theories and Majorana-Yukawa
couplings on the lattice}

\author{\speaker{Yuji Igarashi}%
	 \\
        Faculty\ of\ Education, 
 Niigata\ University, Ikarashi,\\
950-2184, Niigata, Japan\\
        E-mail: \email{igarashi@ed.niigata-u.ac.jp}}

\author{Jan M. Pawlowski\\
Institut f\"ur Theoretische Physik, Universit\"at Heidelberg,
	Philosophenweg 16,\\
	69120 Heidelberg, Germany\\ 
E-mail: \email{j.pawlowski@thphys.uni-heidelberg.de} }

\abstract{
  The construction of  CP-invariant lattice chiral gauge
  theories and the construction of lattice Majorana fermions with chiral Yukawa couplings 
  is subject to topological obstructions. In the present work we suggest lattice extensions of
  charge and parity transformation for Weyl fermions. This enables us 
  to construct lattice chiral gauge theories that are CP invariant. For the construction of 
  Majorana-Yukawa couplings, we discuss two models
  with symplectic Majorana fermions: a model with  
  two symplectic doublets, and one with an
  auxiliary doublet.   
}

\FullConference{The XXVII International Symposium on Lattice Field Theory - LAT2009\\
		 July 26-31 2009\\
		 Peking University, Beijing, China}

\begin{document}

\section{Introduction}

Despite the considerable success in the formulation of chiral symmetry
on the lattice
\cite{Ginsparg:1981bj,Neuberger:1997fp,Hasenfratz:1998ri,Luscher:1998pq}
based on the Ginsparg-Wilson relation, there remain several unsolved
problems. One of them concerns the construction of CP
invariant chiral gauge theories on the lattice \cite{Fujikawa:2002fis,
  Hasenfratz:2005ch}.  Another problem concerns the definition of
Majorana fermions in the presence of Yukawa
couplings \cite{Fujikawa:2002fi,Suzuki:2004ht}.  These problems are
closely related to the requirements of locality and of avoiding
species doublers, which are basic issues for chiral symmetry on the
lattice. Loosely speaking, the above problems relate to the fact
that chiral symmetry for a lattice Dirac action $\bar\psi D \psi$ with
Ginsparg Wilson Dirac operator $D$ requires an asymmetric treatment of
$\psi$ and $\bar\psi$. In turn, CP symmetry and Majorana-Yukawa
couplings require a symmetric treatment of $\psi$ and $\bar\psi$.

In this paper, we put forward possible solutions to these problems. We
first discuss the obstructions in constructing lattice chiral gauge
theories with CP invariance.  Due to the Nielsen-Ninomiya no-go
theorem
\cite{Karsten:1980wd,Nielsen:1980rz,Karsten:1981gd,Friedan:nk},
consistent chiral projection operators necessarily depend on the Dirac
operator, see e.g. \cite{Jahn:2002kg}. It is natural
to assume that the modified chiral symmetry on the lattice induces
modifications of charge and parity transformations on the lattice. Here we
define lattice extensions of charge and parity transformations for
Weyl fermions \cite{Igarashi:2009yj} that explicitly depend on the
chiral projection operators. This will be legitimate because CP is a
discrete symmetry, and enables us to show CP symmetry in chiral gauge
theories.  We then construct Majorana-Yukawa actions by employing symplectic
Majorana fermions. In addition to the model with two symplectic
doublets discussed in \cite{Igarashi:2009yj, Pawlowski:2007ua}, we
also construct a model with an auxiliary symplectic doublet following the
idea given in \cite{Luscher:1998pq,Kikukawa:20005}.

\section{Obstructions in showing CP invariance of chiral gauge theory} 
Let us consider the lattice action of a chiral gauge theory,
\begin{eqnarray}
S_{CGT} =
  \sum_{x,y \in \Lambda} \bar\psi(x) \Bigl(\frac{1 -
    \gamma_{5}}{2}\Bigr) D(U)~ (x-y) \Bigl(\frac{1 +
    \hat\gamma_{5}}{2}\Bigr){\psi}(y) 
\label{eq:CGT-action}
\,, 
\end{eqnarray} 
where the Dirac operator $D(U)$ with link variables $U$ is used to
define ${\hat\gamma}_{5}=\gamma_{5}(1-a D(U))$. With these definitions
the GW relation reads,
\begin{eqnarray}
 \gamma_{5} D(U) +  D(U) {\hat\gamma}_{5}  = 0\,.
\end{eqnarray} 
The standard CP transformation is an operation
\begin{eqnarray}
\psi \to - W^{-1}~ \bar\psi^{T}, \qquad
\bar\psi \to \psi^{T} ~W
\label{eq:CP-tra1}
\,,
\end{eqnarray} 
where $W=CP$ is product of the parity $P = \gamma_{4}$ and the charge
conjugation matrix $C$ satisfies 
\begin{eqnarray}
C \gamma_{\mu}~C^{-1} = -\gamma_{\mu}^{T}, \quad C~C^{\dagger}=1, \quad
 C = - C^{T} \,.
\end{eqnarray} 
The CP transformation \eq{eq:CP-tra1} is not an 
invariance of the action \eq{eq:CGT-action}, simply because
\begin{eqnarray}
W \gamma_{5}~W^{-1} = -\gamma_{5}^{T} \neq - \hat\gamma_{5}^{T}
\,.
\end{eqnarray} 
We observe, however, that the action \eq{eq:CGT-action} would be CP
invariant with the standard parity transformation if the charge
conjugation maps $\gamma_{5}$ to $\hat\gamma_{5}^{T}$. Therefore, one
may construct a lattice extension of the charge conjugation
matrix, $\hat{C}$, which satisfies
\begin{eqnarray}
\hat{C} \gamma_{5}~\hat{C}^{-1} = \hat\gamma_{5}^{T}\,.
\label{eq:C-hat}
\end{eqnarray} 
A solution to this equation is given by $\hat{C}= C~(1-aD/2)$, which
vanishes for $D = 2/a$.  Indeed, any attempt of constructing smooth
mappings between two different types of chiral projection operators 
$P = (1+\gamma_{5})/2$ and $\hat{P}= (1+ \hat\gamma_{5})/2$ fails. 
This can be understood as follows: For general chiral projection
operators $P$ and $\hat{P}$ satisfying 
\begin{eqnarray}\label{eq:genchiral} 
(1-P)\, D = D\, \hat P\,, 
\end{eqnarray}
it has been shown in \cite{Jahn:2002kg} 
that the projection operators
$P$ and $\hat P$ carry a winding number that is related to the
total chirality $\chi$ of the system at hand,
 \begin{eqnarray}\label{eq:chirality} 
\chi  = n[\hat P]-n[1-P]\,, \qquad {\rm with} \qquad 
n[P] \equiv \frac{1}{{2}!}\left(\frac{i}{2\pi}\right)^{2} 
\int_{T^{4}} \tr\,P (\dd P)^{4} \,.
\in \Z \,.
\end{eqnarray} 
\Eq{eq:chirality} therefore entails that for odd total chirality,
e.g.\ a single Weyl fermion, $\hat P\psi$ and $\bar\psi~P$ live in
topologically different spaces. Hence there are no smooth mappings
connecting them. Note that this theorem applies to a wide class of
Dirac operators including Ginsparg-Wilson Dirac operators as a special
case.

\section{Lattice extension of C and/or P transformation}  
The absence of smooth mappings between the two spaces specified by
$\hat P\psi$ and $\bar\psi~P$ may not be a problem, because CP is a
discrete symmetry. We conclude that we simply have to include the
chiral projection operators in the definition of C and/or P
transformation. Consequently we define a lattice extension of
charge conjugation for Weyl fermions \cite{Igarashi:2009yj}:
\begin{eqnarray}
&&\bar\psi(x) \Bigl(\frac{1 \pm \gamma_{5}}{2}\Bigr) \to \sum_{y \in \Lambda} 
\psi^{T}(y) C\Bigl( 
\frac{1 \pm \tilde\gamma_{5}(U^{C})}{2}\Bigr)(y, x)\nn\\
&& \sum_{y \in \Lambda}\Bigl(\frac{1 \pm \hat\gamma_{5}(U)}{2}\Bigr)(x, y)\psi(y)
 \to - 
\Bigl(\frac{1 \pm \gamma_{5}}{2}\Bigr)C^{-1} \bar\psi^{T}(x)\nn\,,
\end{eqnarray}
where $\tilde\gamma_{5}= (1-a D(U))\gamma_{5}$.
For link variable $U$, we use the standard C and P transformations:   
\begin{eqnarray}
U_{\mu}(x) \to U_{\mu}^{C}(x) &=& \bigl(U_{\mu}^{\dagger}\bigr)^{T}(x) \nn\\
U_{\mu}(x) \to U_{\mu}^{P}(x) &=& 
\left\{
		\begin{array}{ll}
		 U_{i}^{\dagger}(x_{P}- a {\hat i}) & for~~  i=1,2,3. \\[1ex]
		 U_{4}(x_{P}) & {} 
		\end{array}
               \right.
\label{eq:new-C}
\end{eqnarray}
We also use the standard parity transformation for spinors
\begin{eqnarray} 
\psi(x) \to \psi^{P}(x) = P^{-1} \psi(x_{P}), \qquad 
\bar\psi(x) \to \bar\psi^{P}(x) = \bar\psi(x_{P})P
\label{eq:P-tr} \,.
\end{eqnarray}
Using \eq{eq:new-C} and \eq{eq:P-tr}, 
we obtain the relations
\begin{eqnarray} 
C {D}(U^{C})  C^{-1} &=& \bigl(D(U)\bigr)^{T}\nn\\
P D(U^{P}) P^{-1} (x,y) &=& D(U)(x_{P}, y_{P})
 \,.             
\end{eqnarray}
It is straightforward to show CP invariance of \eq{eq:CGT-action}, 
\begin{eqnarray} 
&& \bar\psi(1 - \gamma_{5}) D(U)(1 +\hat\gamma_{5})\psi \nn\\
&& \stackrel{C}\longrightarrow
 -\psi^{T}C(1-\tilde\gamma_{5}(U^{C}))D(U^{C})(1+\gamma_{5})
C^{-1}\bar\psi^{T} \nn\\ 
&&\quad  = \bar\psi(1+ \gamma_{5})D(U)(1-\hat\gamma_{5}(U))\psi  \nn\\
&& \stackrel{P}\longrightarrow
 \bar\psi(1 - \gamma_{5}) D(U)(1 +\hat\gamma_{5}(U))\psi \,.
\end{eqnarray}
Some remarks are in order:

\begin{itemize} 
\item[(1)] Performing the charge conjugation \eq{eq:new-C} twice, one finds
\begin{eqnarray} 
\bar\psi (1+\gamma_5)/2 \to \bar\psi (1+\gamma_5)/2, \quad 
(1+\hat\gamma_5)\psi/2 \to (1+\hat\gamma_5)\psi/2, 
\end{eqnarray}
as it should be. 

\item[(2)] In the continuum limit, the charge conjugation \eq{eq:new-C} tends towards the standard one, 
\begin{eqnarray} 
\psi \to - C^{-1}~ \bar\psi^{T} \quad \bar\psi \to \psi^{T} ~C\,.
\end{eqnarray}

\item[(3)] For the functional measure introduced by L$\ddot{\rm u}$scher,
\begin{eqnarray}
&&{\cal D}\psi {\cal D}\bar\psi = \prod_{j} (d c_{j} d {\bar c}_{j})\nn\\
&& \psi(x) = \sum_{j} v_{j}(x) c_{j},\quad \bar\psi(x) = \sum_{j}{\bar v}_{j}{\bar c}_{j}\nn\\
&& \left(\frac{1 + \hat\gamma_{5}}{2}\right)v_{j} = v_{j}, \quad {\bar v}_{j} \left(\frac{1 -\gamma_{5}}{2}\right)= {\bar v}_{j}\,,
\end{eqnarray}
a CP transformation with \eq{eq:new-C} acts as $c_{j} \Leftrightarrow \bar{c}_{j}$,
and therefore ${\cal D}\psi {\cal D}\bar\psi$ remains invariant.
\end{itemize} 

\noindent In the construction put forward above we have modified the charge
conjugation. Alternatively we could use the standard charge
conjugation while modifying the parity transformations for Weyl fermions:
\begin{eqnarray}
&& \left(\frac{1\pm \gamma_{5}}{2}\right) \psi(x) \to P^{-1} 
\sum_{y \in \Lambda} \left(\frac{1\mp \hat\gamma_{5}(U)}{2}\right)(x_{P},y_{P}) \psi(y_{P}) \nn\\
&& \sum_{y \in \Lambda} \bar\psi(y) \left( \frac{1\pm \tilde\gamma_{5}(U)}{2}\right)(y,x) 
\to \bar\psi(x)\left( \frac{1\mp \gamma_{5}}{2}\right)  P\,.
\label{eq:modP}\end{eqnarray} 
The modified CP-transformations with \eq{eq:modP} are an invariance of the theory.

\section{Majorana fermions and Yukawa couplings}
Majorana spinors are defined by imposing a reality condition
with $B=\gamma_{5}C$,
\begin{eqnarray}
\psi^{*}= B \psi \qquad \Rightarrow \qquad \psi^{* *}= B^{*} B \psi\,. 
\end{eqnarray} 
For four dimensional Euclidean space, we have $B^{*} B =-1$ which
leads to $\psi^{* *} =~-\psi$. Therefore, there are no Majorana
spinors satisfying the reality constraint \cite{Nicolai:1978,van
  Nieuwenhuizen:1996}. This difficulty can be circumvented by doubling
the fermions and implementing a symplectic Majorana condition
\begin{eqnarray}
&& \psi_{1}^{*} = B~\psi_{2} , \quad \psi_{2}^{*}=-B~ \psi_{1} \nn\\
&&\Rightarrow \psi_{a}^{**} = \epsilon_{ab}B^{*}\psi_{b}^{*}= \epsilon_{ab}
\epsilon_{bc} B^{*}B \psi_{c} = \psi_{a}\quad (a,b=1,2)\,.
\end{eqnarray} 
On the lattice, a further doubling of degrees of freedom is 
needed for a chirally invariant theory with GW Dirac operator
$D$ \cite{Igarashi:2009yj,Pawlowski:2007ua}. 
Introducing symplectic pairs of Majorana spinors, $(\Psi_{1}, \Psi_{2})$
and $(\psi_{1}, \psi_{2})$, we construct a chiral Yukawa theory: 
\begin{eqnarray}
S &=& S_{0} + S_{Y}\nn\\
S_{0} &=& 
\sum \left({\psi}_{1}^{T}C {D} \Psi_{1} + {\psi}_{2}^{T}C {
      D} \Psi_{2} \right)\nn\\
S_{Y} &=& \frac{g}{4} \sum \biggl[\Bigl\{\psi_{1}^{T}C (1+ \gamma_{5})\varphi 
(1+ \hat\gamma_{5})\Psi_{1}\nn\\
&&  +\psi_{1}^{T}C (1- \gamma_{5})\varphi^{*} 
(1- \hat\gamma_{5})\Psi_{1}\Bigr\} + \Bigl\{ 1 \to 2 \Bigr\} \biggr] \,.
\end{eqnarray} 
The action is invariant under the chiral transformations 
\begin{eqnarray}
\delta \psi_{1} &=& i \varepsilon \gamma_{5} \psi_{1},\quad 
\delta \psi_{2} = - i \varepsilon \gamma_{5} \psi_{2} \nn\\
\delta \Psi_{1} &=& i \varepsilon \hat\gamma_{5} \Psi_{1},\quad 
\delta \Psi_{2} = -i \varepsilon \hat\gamma_{5} \Psi_{2}\nn\\
\delta \varphi &=& 2i \varepsilon  \varphi\,.
\end{eqnarray} 
In the above we have introduced two symplectic doublets. It is possible
to make one of them an auxiliary doublet. Following 
\cite{Luscher:1998pq,Kikukawa:20005}, we consider a free field action of two  
symplectic Majorana doublets,
\begin{eqnarray}
S_{0} = \sum \biggl[\Bigl(\psi_{1}^{T}C D\psi_{1}-
 \frac{2}{a}\Psi_{1}^{T}C \Psi_{1}\Bigr) + \Bigl(1 \to 2\Bigr)
\biggr]\,,
\label{eq:action2}
\end{eqnarray} 
where $(\Psi_{1},\Psi_{2})$ are auxiliary fields. The action 
\eq{eq:action2} is invariant under the symmetric chiral transformations 
\begin{eqnarray}
&& \delta \psi_{1} =i \varepsilon
 \gamma_{5}\left(1-\frac{a}{2}D\right)\psi_{1} + i \varepsilon
 \gamma_{5} \Psi_{1},\quad~~
\delta \Psi_{1} = i \varepsilon \gamma_{5}\frac{a}{2}D\psi_{1}
\nn\\
&& \delta \psi_{2} = -i \varepsilon
 \gamma_{5}\left(1-\frac{a}{2}D\right)\psi_{2} - i \varepsilon
 \gamma_{5} \Psi_{2},\quad
\delta \Psi_{2} = - i \varepsilon \gamma_{5}\frac{a}{2}D\psi_{2}\nn \\
&& \delta \varphi = 2i \varepsilon \varphi\,.
\label{eq:chiral-tr2}
\end{eqnarray}
Since
\begin{eqnarray}
\delta (\psi_{1} + \Psi_{1}) = i \varepsilon
 \gamma_{5}(\psi_{1} + \Psi_{1}),~~
\delta (\psi_{2} + \Psi_{2}) =- i \varepsilon
 \gamma_{5}(\psi_{2} + \Psi_{2})\,,
\end{eqnarray}
$\psi_{a}+\Psi_{a}$ can be used to construct a chirally invariant Yukawa coupling,
\begin{eqnarray}
S_{Y} &=& \frac{g}{2}\sum \biggl[\Bigl\{(\psi_{1}+ \Psi_{1})^{T}C \varphi
 (1+\gamma_{5}) (\psi_{1}+ \Psi_{1})\nn\\
&& + (\psi_{2}+ \Psi_{2})^{T}C \varphi^{*}
 (1-\gamma_{5}) (\psi_{2}+ \Psi_{2})
\Bigr\} + \Bigl\{1 \to 2\Bigr\}
\biggr]\,.
\end{eqnarray}
It is easy to see that the total action $S_{0} + S_{Y}$ is invariant
under \eq{eq:chiral-tr2}. 

\section{Discussion and summary}

We have discussed the construction of CP-invariant chiral gauge
theories, as well as that of CP-invariant Majorana-Yukawa couplings on
the lattice. Both problems are closely related to the fact that chiral
projection operators on the lattice necessarily depend on the Dirac
operator.  This already suggests to include the chiral projection
operators explicitly in the definition of charge conjugation and/or parity
transformation for the Weyl fermions on the lattice. On the basis of
these modified transformations we have constructed CP-invariant
actions.  We have also shown that the C and P transformations tend
toward the standard C and P transformations in the continuum limit and
leave the path integral measure invariant.

Let us also discuss some other approaches to the CP problem. In
the interesting work \cite{Hasenfratz:2007dp,Gattringer:2008je}, chiral projection
operators are constructed which are independent of gauge fields. Then 
CP invariance of chiral gauge theory is shown by using an 8-component
notation. For this work, one has to examine whether the formulation gives the
correct fermionic degrees of freedom in 4-component notation.  In
\cite{Cundy:2009ab}, a renormalization group approach is discussed to
give a symmetric form of chiral projection operators. There, locality
of those operators has to be examined.

For the construction of the Majorana-Yukawa couplings, we have built a
model with an auxiliary doublet of symplectic Majorana fermions in
addition to the two doublet model discussed in \cite{Igarashi:2009yj}. An
extension of the formalism based on the symplectic Majorana condition
to supersymmetric theory on the lattice will be interesting and
challenging.

\noindent {\bf Acknowledgements}

YI would like to thank the organisers of {\it Lattice 2009} for all
their efforts which made this inspiring conference possible.  He would
like to thank the Institute of Theoretical Physics in Heidelberg for
hospitality.  We are also grateful to F.~Bruckmann and N.~Cundy for useful
discussions.

\end{document}